\RequirePackage{fix-cm}
\documentclass[smallextended]{svjour3}       
\smartqed  
\usepackage{graphicx}
\usepackage{mathptmx}      
\newcommand{\1}{\mbox{1}\hspace{-0.25em}\mbox{l}}
\begin{document}

\title{Energy dependence of $K^-$-``$pp$'' effective potential 
derived from coupled-channel Green's function
}


\author{Takahisa Koike          \and
        Toru  Harada 
}


\institute{T. Koike \at
Strangeness Nuclear Physics Laboratory, 
RIKEN Nishina Center, Wako-shi, Saitama 351-0198, Japan \\
              \emph{tkoike@riken.jp}           
           \and
           T. Harada \at
Research Center for Physics and Mathematics,
Osaka Electro-Communication University, Neyagawa, Osaka 572-8530, Japan
}

\date{Received: date / Accepted: date}

\maketitle

\begin{abstract}
We investigate the energy dependence of 
a single-channel effective potential between the $K^-$ and 
the ``$pp$''-core nucleus, which can be obtained 
as an $K^-$-``$pp$'' equivalent local potential 
from a coupled-channel model for $\bar{K}(NN)$-$\pi(\Sigma N)$ 
systems.
It turns out that the imaginary part of the resultant 
potential near the $\pi \Sigma N$ decay threshold can 
well approximate the phase space suppression factor of 
$K^-pp \to \pi \Sigma N$ decay modes. 
The effects on the pole position of the 
$\pi(\Sigma N)$ state 
in the $\pi \Sigma N$ channel are also discussed.
\end{abstract}

\section{Introduction}
\label{intro}

In order to elucidate the nature of antikaon-nucleon ($\bar{K}N$) 
interaction in high density nuclear matter, it is an important 
subject to clarify whether the ``deeply-bound kaonic nucleus''
exists or not. In particular,  the
$[{\bar K}\otimes \{NN \}_{I=1}]_{I=1/2}$, $J^\pi=0^-$ bound state, 
which is called ``$K^-pp$'' here, is suggested to 
be the lightest and most fundamental kaonic nucleus~\cite{YA02}.
A new experimental search of $K^-pp$ via the 
$^3$He(in-flight $K^-$, $n$) reaction has been planned at J-PARC
as E15 experiment~\cite{Iwa06}.
We theoretically have discussed the expected inclusive and 
semi-exclusive spectra for the $^3$He(in-flight $K^-$, $n$) reaction
within the framework of the distorted-wave impulse approximation (DWIA) .
We have employed a phenomenological single-channel $K^-$-``$pp$'' (complex) 
effective potential between the $K^-$ and 
the ``$pp$''-core nucleus~\cite{Koi07,Koi09}, which has the form of 
\begin{equation}
U^{\rm eff}(E; r) = 
\left( V_0 + i \, W_0 \, f(E) \,  \right) \exp[-(r/b)^2], 
\label{eq1}
\end{equation}
where $V_0$ and $W_0$ are strength parameters in real and imaginary parts,
respectively, and $b$ is the range parameter. 
$f(E)$ is the phase space suppression factor of the $K^-pp \to \pi \Sigma N$ decay 
modes~\cite{Mar05} where $E$ is the energy measured from the 
$K^-pp$ threshold. 

In the single-channel framework,
we have shown that behavior of the $^3$He(in-flight $K^-$, $n$) spectrum 
can be understood in the ``moving pole'' picture~\cite{Koi09};
a pole of the $K^-pp$ bound state moves in the complex energy plane 
as a function of the energy $E$ on the real axis
because the $K^-$-``$pp$'' effective potential is 
considerably energy-dependent.
A trajectory of its moving pole governs the shape of the spectrum. 
Thus, the validity of our calculations relies in part on 
whether the energy dependence of Eq.(\ref{eq1}) is appropriate near the 
$\pi \Sigma N$ decay threshold or not.
To examine this subject, we have extended the previous 
$K^-pp$ single channel description to 
the  $\bar{K}(NN)$-$\pi(\Sigma N)$  coupled-channel (CC)  description 
because the energy dependence of Eq.(\ref{eq1}) should originate 
from eliminating the $\pi \Sigma N$ channel 
in such a CC scheme.

In this article,
we investigate a single-channel $K^-$-``$pp$'' equivalent local potential 
which is derived from CC Green's functions for 
$K^-(pp)$-$\pi(\Sigma N)$ systems, and evaluate the energy dependence 
of this effective potential to be compared with a phenomenological 
one~\cite{Koi09} which is determined within Eq.(\ref{eq1}).

\section{Coupled-channel model}

Let us consider the following CC Green's function $G_{ij}$
for $K^-(pp)$-$\pi(\Sigma N)$ systems;
\begin{eqnarray}
\left[
\begin{array}{cc}
E_1-T^{(l)}_1-U_{11}(r) & -U_{12}(r)  \\
-U_{21}(r) & E_2-T^{(l)}_2-U_{22}(r)  \\
\end{array}
\right]
\left[
\begin{array}{cc}
G^{(l)}_{11}(r,r') & G^{(l)}_{12}(r,r')  \\
G^{(l)}_{21}(r,r') & G^{(l)}_{22}(r,r')  \\
\end{array}
\right]
= \delta(r'-r) \1,
\label{eq2}
\end{eqnarray}
where the channel 1 (2) refers to the $K^-$ and $pp$-core system 
(the $\pi$ and $\Sigma N$-core system):
$T^{(l)}_{1,2}$ denote the kinetic energies for 1, 2, and the energy 
$E_2 = E_1  - E_{\rm th}(\pi \Sigma N)$ where 
$E_{\rm th}(\pi \Sigma N)$ $\simeq -100$ MeV below the $K^-pp$ threshold.
$U_{11}$, $U_{22}$ and $U_{12}(=U_{21})$ denote 
the diagonal and coupling potentials, respectively.
For simplicity, we assume an energy-independent Gaussian form as 
\begin{equation}
\left[
\begin{array}{cc}
U_{11}(r) & U_{12}(r)  \\
U_{21}(r) & U_{22}(r)  \\
\end{array}
\right]
= 
\left[
\begin{array}{cc}
V_1 + i \, W_1 & V_{\rm c} \\
V_{\rm c} & V_2 + i \, W_2 \\
\end{array}
\right]
\exp[-(r/b)^2], 
\label{eq3}
\end{equation}
where 
$V_{1,2}$ and $V_{\rm c}$ denote the strength parameters of the diagonal 
and coupling potentials, respectively, and 
$W_{1,2}$ describe effects of the other decay modes.
These parameters in Eq.(\ref{eq3}) should be determined 
so as to reproduce the proper values of the binding energy and width 
which are obtained by single-channel calculations with 
$U^{\rm eff}$ in Eq.(\ref{eq1}).
If we use CC Green's functions in Eq.(\ref{eq2}), 
we expect to confirm the previous results~\cite{Koi07,Koi09}
for the $^3{\rm He}$(in-flight $K^-$, n) spectrum
within the DWIA calculation.

Now we define the $K^-$-``$pp$'' equivalent local potential
for channel 1, $\tilde{U}_{11}^{\rm eff}$ as
\begin{equation}
\{ E_1-T^{(l)}_1-\tilde{U}_{11}^{\rm eff}(E;r) \} \, 
G^{(l)}_{11}(E;r,r')  = \delta(r'-r),
\label{eq4}
\end{equation}
where $G^{(l)}_{11}$ is the (1,1) component of a solution 
of Eq.(\ref{eq2}). Then, we get 
\begin{equation}
\tilde{U}_{11}^{\rm eff}(E;r) \, G^{(l)}_{11}(E;r,r')
= U_{11}(r) \, G^{(l)}_{11}(E;r,r') 
+ U_{12}(r) \, G^{(l)}_{21}(E;r,r'). 
\label{eq5}
\end{equation}
By multiplying the initial wave function $\phi_1(r')$ 
as a bound state in channel 1, and by integrating over $r'$, 
we obtain the expression of the equivalent local potential as 
\begin{equation}
\tilde{U}_{11}^{\rm eff}(E;r) \,
= U_{11}(r) +  U_{12}(r) \,
\displaystyle 
\frac{\displaystyle \int_0^{\infty} G^{(l)}_{21}(E;r,r') \, \phi_1(r') \, dr'}
{\displaystyle  \int_0^{\infty} G^{(l)}_{11}(E;r,r') \, \phi_1(r') \,  dr'},
\label{eq6}
\end{equation}
which is defined under the 
boundary condition at {\it every} $E$ on the physical axis,  
because Green's functions can be calculated numerically at every points
in the complex energy plane.
This is the advantage that we used Green's functions rather than 
wave functions.

\begin{figure}[t]
\begin{center}
\includegraphics[angle=90,width=0.98\linewidth]{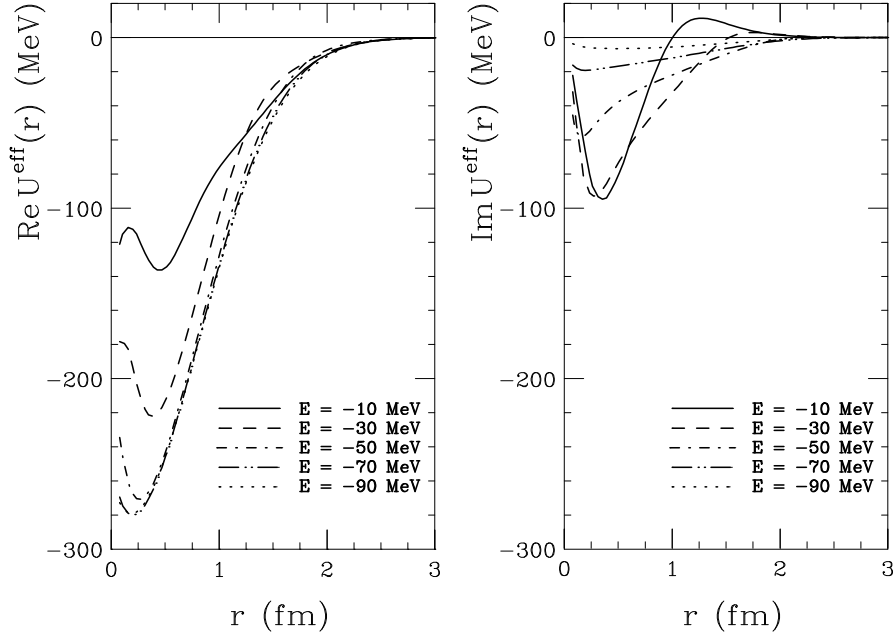}
\end{center}
\caption{Real (left) and imaginary (right)  parts of the equivalent 
local potential $\tilde{U}_{11}^{\rm eff}$ for the $K^-$-``$pp$'' channel. 
Potential parameters of Eq.(\ref{eq3}) are used as 
$V_1 = -300$ MeV, $V_2 = -150$ MeV, 
$V_{\rm c} = -100$ MeV, $W_1=W_2=0$ MeV and $b = 1.09$ fm.
The energy $E$, which is measured from the $K^-pp$ threshold, 
is varied form $-10$ MeV to $-90$ MeV.}
\end{figure}

\section{Results and discussion}

Figure 1 shows the calculated equivalent local potential
$\tilde{U}_{11}^{\rm eff}$ 
with $V_1 = -300$ MeV, $V_2 = -150$ MeV, 
$V_{\rm c} = -100$ MeV, $W_1=W_2=0$ MeV and $b = 1.09$ fm, 
in which the potential parameters are determined so that the binding 
energy and width of $K^-pp$ bound state are consistent with 
those calculated by Yamazaki and Akaishi~\cite{YA02}.
We find that the strength of the imaginary part of 
$\tilde{U}_{11}^{\rm eff}$ becomes shallower 
as $E$ goes to $E_{\rm th}(\pi \Sigma N)$.
This behavior is approximately equivalent to that of the energy dependence 
obtained from the phase space factor in Eq.(\ref{eq1}), whereas 
the shape of $\tilde{U}_{11}^{\rm eff}$ is not a Gaussian function exactly.
Therefore, we estimate the potential strength of
$(V^{\rm eff}(E), W^{\rm eff}(E))$ 
which is obtained with the help of the volume integrals;
\begin{equation}
V^{\rm eff}(E) + i \, W^{\rm eff}(E)= 
{\displaystyle \int_0^{\infty} \tilde{U}_{11}^{\rm eff}(E;r) \, r^2 dr}
\bigg/  {\displaystyle  \int_0^{\infty} \exp[-(r/b)^2] \, r^2 dr}.
\label{eq7}
\end{equation}
Figure~2 shows values of $W^{\rm eff}(E)$ as a function of the energy $E$, 
together with the fitted curve of $W_0 \times f(E)$ in Eq.(\ref{eq1}). 
The results are as follows:
\begin{itemize}
\item
If no bound state exists in the $\pi(\Sigma N)$ channel ($V_2 =-150$ MeV),
Im $\tilde{U}_{11}^{\rm eff}$
approximates to the phase space factor $f(E)$ that is used 
in Eq.(\ref{eq1}) near the $K^-pp \to \pi \Sigma N$ decay 
threshold.
 
\item
If a bound state exists in the $\pi(\Sigma N)$ channel ($V_2 =-300$ MeV),
the energy dependence of 
Im $\tilde{U}_{11}^{\rm eff}$
considerably differs from that of Eq.(\ref{eq1}) 
due to modification of the phase volume via a pole which is located 
near the $\pi\Sigma N$ threshold.
\end{itemize}
In summary, we have investigated the energy dependence of 
the single-channel $K^-$-$``pp"$ equivalent local potential $\tilde{U}_{11}^{\rm eff}$ 
derived from the $\bar{K}(NN)$-$\pi(\Sigma N)$ model.
It has turned out that the imaginary part of $\tilde{U}_{11}^{\rm eff}$ 
near the $\pi \Sigma N$ decay threshold can 
well approximate the phase space suppression factor of 
$K^-pp \to \pi \Sigma N$ decay modes. 
If potential parameters in Eq.(\ref{eq3}) are replaced by the energy-dependent ones,
as obtained from chiral $\bar{K}N$-$\pi Y$ dynamics \cite{Jid03},  
their energy dependence would additionally contribute to the effective potential.
The further detailed investigation is now in progress.

\begin{figure}[t]
\begin{center}
\vspace{1mm}
\includegraphics[angle=90,width=0.95\linewidth]{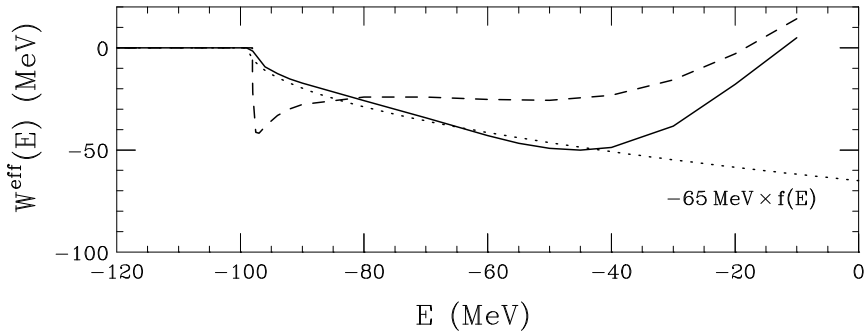}
\end{center}
\caption{
Energy dependence of the imaginary part $W^{\rm eff}(E)$ of 
the $K^-$-$(pp)$ equivalent local potential $\tilde{U}_{11}^{\rm eff}$.
The solid and dashed lines denote the cases with 
$V_2 =-150$ MeV and $-300$ MeV in Eq.(\ref{eq3}), respectively. 
The dotted line denotes values obtained by $W_0 \times f(E)$ 
with $W_0=-65$ MeV. 
}
\end{figure}

\end{document}